\documentclass[twocolumn]{aastex62}

\usepackage{epsfig}
\usepackage{epstopdf}
\usepackage{graphics}
\usepackage{amsmath}
\usepackage{xcolor}
\usepackage{aas_macros}

\usepackage{amsmath,amssymb,amsfonts,amsbsy}
\usepackage{mathrsfs}
\usepackage{url}
\usepackage{multirow}
\usepackage{xcolor}
\usepackage{color}
\usepackage{ulem}
\usepackage{enumerate}

\renewcommand{\figurename}{Fig.}
\renewcommand{\tablename}{Tab.}
\begin{document}

\title{Electron and positron spectra in the three dimensional spatial-dependent propagation model}

\correspondingauthor{Wei Liu, Yi-Qing Guo}
\email{liuwei@ihep.ac.cn, guoyq@ihep.ac.cn}

\author{Zhen Tian}
\affiliation{Institute of Applied Physics and Computational Mathematics, Chinese Academy of Engineering, Beijing 100094,China}

\author{Wei Liu}
\affiliation{Key Laboratory of Particle Astrophysics,
Institute of High Energy Physics, Chinese Academy of Sciences, Beijing 100049, China
}

\author{Bo Yang}
\affiliation{Institute of Applied Physics and Computational Mathematics, Chinese Academy of Engineering, Beijing 100094,China}

\author{Xue-Dong Fu}
\affiliation{Institute of Applied Physics and Computational Mathematics, Chinese Academy of Engineering, Beijing 100094,China}

\author{Hai-Bo Xu}
\affiliation{Institute of Applied Physics and Computational Mathematics, Chinese Academy of Engineering, Beijing 100094,China}

\author{Yu-hua Yao}
\affiliation{College of Physical Science and Technology, Sichuan University, Chengdu, Sichuan 610064, China
}
\affiliation{Key Laboratory of Particle Astrophysics,
Institute of High Energy Physics, Chinese Academy of Sciences, Beijing 100049, China
}

\author{Yi-Qing Guo}
\affiliation{Key Laboratory of Particle Astrophysics,
Institute of High Energy Physics, Chinese Academy of Sciences, Beijing 100049, China
}

%diverse
% study
%apply spatial-dependent propagation plus a local source model
%apply to 
\begin{abstract}
The spatial-dependent propagation model has been successfully used to explain diverse observational phenomena, including the spectral hardening of cosmic-ray nuclei above $200$ GV, the large-scale dipole anisotropy and the diffusive gamma distribution. In this work, we further apply the spatial-dependent propagation model to both electrons and positrons. To account for the excess of positrons above $10$ GeV, an additional local source is introduced. And we also consider a more realistic spiral distribution of background sources. We find that due to the gradual hardening above $10$ GeV, the hardening of electron spectrum above tens of GeV can be explained in the SDP model and both positron and electron spectra less than TeV energies could be naturally described. The spatial-dependent propagation with spiral-distributed sources could conforms with the total electron  spectrum in the whole energy. Meanwhile compared with the conventional model, the spatial-dependent propagation with spiral-distributed sources could produce larger background positron flux, so that the multiplier of background positron flux is $1.42$, which is much smaller than the required value by the conventional model. Thus the shortage of background positron flux could be solved. Furthermore we compute the anisotropy of electron under spatial-dependent propagation model, which is well below the observational limit of Fermi-LAT experiment.
\end{abstract}

% Moreover a local pulsar is introduced to account for the hardening of total electron spectrum and the knee-like structure. Under this scenario, the spectra of both electron and positron could be well described.
% better compute the contribution background of cosmic ray electron and positron, the background sources are considered to distribute along the spiral arms instead of the common . 
 %However, it is hard to explain by the standard model, whose CRE spectrum performs softer as the
 %increasing of energy due to the severe energy losses by synchrotron radiation and inverse Compton
 %scattering processes during propagation.
 % in which two improvements are considered: one is that
 %and the other is that a local pulsar contribute to the hardening and knee-like structures. 


%(including both electrons and positrons)
%But subject to the finite collection area, For the electron flux,  the precise space-borne observations hardly extend beyond $1$ TeV.
%for TeV energies, the ground-based experiments have to be applied and more interesting structures are uncovered thanks to these improved technology. 
%During the past years, great progresses in the measurement of cosmic ray electron and positron have been made by the new generations of space-borne and ground-based experiments. 
% the Fermi-LAT collaboration update their measurements on cosmic ray electron (CRE) flux (including both electron and positron) up to $2$ TeV \citep{PhysRevD.95.082007}, while the observations from the DAMPE satellite extend up to $5$ TeV with a higher energy resolution and a low background \citep{chang2008excess}. Both measurements, as well as
%, reveal
%that there is
%, which enable the CRE spectrum hardening
\section{Introduction}
The conventional cosmic-ray propagation model predicts that the observed energy spectrum falls as a featureless power law, i.e. $\propto R^{-\nu -\delta}$, with $\nu$ and $\delta$ being the power indexes of  the injection spectrum and diffusion coefficient respectively. However more and more observations disfavor such simple picture. First of all, a significant excess of positrons above $10$ GeV was discovered by the PAMELA experiment \citep{2009Natur.458..607A}. Before long this anomaly was substantiated by the Fermi-LAT experiment \citep{2009PhRvL.102r1101A}. The recent observations from the AMS-02 collaboration extended the measurements up to $600$ GeV with an unprecedented high precision \citep{2014PhRvL.113l1101A,2014PhRvL.113l1102A}. On the other hand, the comprehensive analysis to the AMS-02 electron data \citep{2015PhLB..749..267L} indicate an additional component above $\sim 50$ GeV, which has been observed by the DAMPE satellite \citep{chang2008excess}. Moreover, the HESS experiment reported that there is a spectral break around $1$ TeV in the total electron spectrum (electron + positron), which resembled the knee region at $\sim 4$ PeV in the spectrum of cosmic ray nuclei \citep{2009A&A...508..561A}. This sign was also observed by other experiments such as  MAGIC, VERITAS and the latest DAMPE \citep{2011ICRC....6...47B, 2015arXiv151001269S, chang2008excess}.
   %Beyond tens of
   %TeV, there are still lack of observations. However, these experiments also
   %indicate a sign of CRE excess above several TeV
   %\citep{2009A&A...508..561A,2015arXiv151001269S}.

The overabundance of positrons has called a lot of attention, which implies the existence of extra primary sources. A number of models have been proposed to explain the PAMELA and AMS-02 observations, which can be either astrophysical, such as local pulsars \citep{1970ApJ...162L.181S, 2001A&A...368.1063Z, 2009PhRvL.103e1101Y, 2009JCAP...01..025H, 2015APh....60....1Y, 2012CEJPh..10....1P, 2015arXiv150908227G} and the hadronic interactions inside SNRs \citep{2009PhRvL.103e1104B, 2009PhRvD..80f3003F, 2009ApJ...700L.170H, 2015ApJ...803L..15T}, or more exotic origins like the dark matter self-annihilation or decay \citep{2008PhRvD..78j3520B, Cirelli:2008pk, 2009PhLB..672..141B, 2009PhRvD..79b3512Y, 2009PhRvL.103c1103B, 2009PhRvD..80b3007Z}. For an extensive introduction of relevant models, one can refer to the reviews \citep{2009MPLA...24.2139H, 2010IJMPD..19.2011F, 2012APh....39....2S,  2012Prama..79.1021C, 2013FrPhy...8..794B} and references therein. Additionally, the $e^+/e^-$ ratio can also be interpreted as the charge-sign dependent particle injection and acceleration \citep{2016PhRvD..94f3006M}. For the drop-off of electron spectrum, it is argued to be caused by the radiation cooling of electrons surrounding SNRs \citep{2009A&A...497...17V} or the threshold interaction during the transport of cosmic ray electrons \citep{2009ApJ...700L.170H, 2010SCPMA..53..842W, 2016arXiv161108384J}. 

%According to their measurements, above $\sim 200$ GeV/nucleon, the fluxes of both elements steadily rise with increasing energies. Later on, such feature is identified by the  exper- iment with a higher significance
%which well explains the proton and helium anomalies
%Usually the CR diffusion is regarded as uniform and isotropic on large scale. Thus, it is only a function of rigidity, which does not vary spatially. But
%distribution of 

The spatial-dependent propagation (SDP) model was first used to describe the spectral hardening of primary cosmic-ray proton and helium above $200$ GV which are observed successively by ATIC-2 \citep{2006astro.ph.12377P, 2009BRASP..73..564P}, CREAM \citep{2010ApJ...714L..89A, 2011ApJ...728..122Y, 2017ApJ...839....5Y} and PAMELA \citep{2011Sci...332...69A} and AMS-02 \citep{2015PhRvL.114q1103A, 2015PhRvL.115u1101A} experiments. This scenario was initially introduced by \cite{2012ApJ...752L..13T} as Two Halo model (THM). In this  model, the whole transport volume is divided into two regions. The Galactic disk and its surrounding area are called the inner halo (IH), in which the diffusion property is influenced by the CR sources. Outside of IH, the diffusion approaches to the traditional assumption, i.e. only rigidity dependent. This extensive region is named as outer halo (OH). To reproduce the high energy excess, the diffusion coefficient within IH has a weaker rigidity dependence on average, compared with OH zone. In addition to the spectral hardening, the SDP model is further applied to solve the puzzles of large-scale anisotropy, diffuse gamma ray distribution and so forth \citep{2012ApJ...752L..13T, 2016ApJ...819...54G, 2018PhRvD..97f3008G, 2018arXiv181209673L, 2019arXiv190100249Q}.

%The spectral hardening brings about different alternatives to the traditional CR theory. Most of them fall into, but not limited to, three categories: the acceleration process \citep{2011ApJ...729L..13O, 2012PhRvL.108h1104M, 2010ApJ...725..184B, 2011PhRvD..84d3002Y, 2017ApJ...835..229K}, the transport effect \citep{2012PhRvL.109f1101B, 2012ApJ...752L..13T, 2012JCAP...01..010B, 2012ApJ...752...68V, 2014A&A...567A..33T, 2014ApJ...782...36E, 2015A&A...583A..95A, 2015PhRvD..92h1301T, 2016ApJ...819...54G, 2016PhRvD..94l3007F, 2016ApJ...819...54G, 2016ChPhC..40a5101J, 1674-1137-42-7-075103, 2018PhRvD..97f3008G}, as well as the origin of local source(s) \citep{2012A&A...544A..92B, 2012MNRAS.421.1209T, 2013MNRAS.435.2532T, 2013A&A...555A..48B, 2015RAA....15...15L, 2015ApJ...803L..15T, 2015ApJ...815L...1T, 2017PhRvD..96b3006L, PhysRevLett.120.041103}. One of the popular scenarios is so-called . 

In this work, we further study the SDP model by applying to the observations of CR electrons and positrons. To account for the excess of positrons above $10$ GeV, a nearby young source is introduced. Meanwhile to better compute the fluxes of background electrons and positrons, we further consider a more realistic spiral distribution of sources. We compare three kinds of transport model: conventional, SDP and SDP plus spiral distribution. We find that under traditional axisymmetric distribution of sources, both conventional and  SDP models could not well explain the spectra of both positron and electron with only a local source, and an extra electron component is needed. But with a spiral distribution, the SDP model could well describe the spectra of electrons, positrons as well as the total. We also find that in this case, the required enhancement of background positron flux is only $1.42$, which is much smaller than the conventional model. We further compute the anisotropy of electron and find that  SDP model predicts a much smaller amplitude of anisotropy.

The rest paper is organized in the following way. In Sec.2, both spatial-dependent propagation model and spiral distribution of sources are presented in detail. Sec.3 gives the calculated results and Sec.4 is reserved for the conclusion.

\section{Model Description}
\subsection{Spatial-dependent propagation}
After escaping into the interstellar space, CRs diffuse within the Galaxy by randomly scattering off magnetic waves and MHD turbulence. Besides diffusion, CRs still experience reacceleration, convection, fragmentation, radioactive decay and energy losses before arriving at earth. The corresponding propagation process could be described by a so-called diffusion equation:
\begin{align}
 \nonumber  \dfrac{\partial \psi( \textbf{r},p,t)}{\partial t} &=
         q(\textbf{r},p,t) + \nabla \cdot (D_{xx} \nabla \psi -V_{c}\psi)\\
 \nonumber   & +   \dfrac{\partial}{ \partial p} p^{2}D_{pp} \dfrac{\partial}{\partial p}\frac{1}{p^{2}}\psi
    -   \dfrac{\partial}{\partial p} \left[\dot{p}\psi - \dfrac{p}{3}(\nabla \cdot V_{c}\psi) \right] \\
         & -  \dfrac{\psi}{\tau_{f}} - \dfrac{\psi}{\tau_{r}} ~.
\end{align}
with $\psi(\textbf{r},p,t)$ the CR density per unit momentum $p$ at position $\textbf{r}$. $\tau_{f}$ and $\tau_{r}$ are the characteristic timescales for fragmentation and radioactive decay respectively. $V_{c}$ is the convention velocity. In the diffusive-reacceleration term, $D_{pp}$ is related to $D_{xx}$ by the formula $D_{pp}D_{xx} = \dfrac{4p^{2}v_{A}^{2}}{3\delta(4-\delta^{2})(4-\delta)}$, in which $v_A$ is the Alfv\'en velocity \citep{1994ApJ...431..705S}.  In this work, we adopt the common diffusion-reacceleration model, which is called DR for short. The diffusive halo is approximated as a flat cylinder with the radius of $R = 20$ kpc. The half-thickness $z_h$ of halo is determined by fitting the B/C ratio. 

The CR sources are distributed in the middle of diffusive halo. $q(\textbf{r},p,t)$ represents the source term of CR particles. The spatial distribution of CR sources is paramterized as 
\begin{equation}
f(r, z) = \left(\dfrac{r}{r_\odot} \right)^\alpha \exp \left[-\dfrac{\beta(r-r_\odot)}{r_\odot} \right] \exp \left(-\dfrac{|z|}{z_s} \right) ~,
\label{eq:source_dis}
\end{equation}
with $r_\odot = 8.5$ kpc, $\alpha = 1.09$, and $\beta = 3.87$ \citep{2015MNRAS.454.1517G} respectively. In the direction perpendicular to the Milky Way, the number of SNRs decays as an exponential function, with a mean value $z_{s} = 100$ pc. To fit the low energy spectra, the injection spectra of proton and electron are assumed to have a broken power-law respectively:
\begin{equation}\label{eq:spectrum_CR}
q^{\rm p}({\cal R})=q^{\rm p}_{0}\left\{
\begin{array}{lll}
&  \left(\dfrac{{\cal R}}{{\cal R}_{\rm br}^{\rm p}} \right)^{\nu_{1}^{\rm p}},   &      {{\cal R} \leq {\cal R}_{\rm br}^{p}}\\
\\
& \left(\dfrac{{\cal R}}{{\cal R}_{\rm br}^{\rm p}} \right)^{\nu_{2}^{\rm p}} \exp \left[-\dfrac{{\cal R}}{{\cal R}_{\rm c}^{\rm p}} \right],  & {{\cal R} >  {\cal R}_{\rm br}^{\rm p}}
\end{array} 
\right.
\end{equation}
and 
\begin{equation}\label{eq:spectrum_CRE}
q^{\rm e}({\cal R}) = q^{\rm e}_{0} \left\{
\begin{array}{lll}
\left(\dfrac{{\cal R}}{{\cal R}_{\rm br}^{\rm e}} \right)^{\nu_{1}^{\rm e}},  &  {{\cal R} \leq {\cal R}_{\rm br}^{\rm e}} \\
\\
\left(\dfrac{{\cal R}}{{\cal R}_{\rm br}^{\rm e}} \right)^{\nu_{2}^{\rm e}} \exp \left[-\dfrac{R}{R_{\rm c}^{\rm e} } \right], & {{\cal R} >  {\cal R}_{\rm br}^{\rm e}} \\
\end{array}
\right.
\end{equation}
where $\nu$ and $q_{0}$ are the spectral index and normalization for proton(electron) respectively, and $R_{\rm c}$ is the cut-off rigidity. 

%the sets of parameters of  and $\nu^{e}$ correspond to the , It is approximated to
%which is identical with the one firstly referred in
In the SDP model, the half-thickness of inner and outer halo are $\xi z_{h}$ and $(1-\xi) z_{h}$ respectively \citep{2012ApJ...752L..13T}. Within the inner halo, the magnitude of diffusion is supposed to have an anti-correlation with the radial distribution of background CR sources. The corresponding diffusion coefficient $D_{xx}$ is thus parameterized as:
\begin{equation}
D_{xx}(r, z, \mathcal R) = D_0 F(r, z) \beta^\eta  \left(\dfrac{\mathcal R}{\mathcal R_0} \right)^{\delta(r, z)} ~.
\end{equation}
Both $F(r,z)$ and $\delta(r, z)$ are anti-correlated with the source density \citep{2018PhRvD..97f3008G}, which are paramterized as
\begin{equation}
F(r,z) =
\begin{cases}
g(r,z) +\left[1-g(r,z) \right] \left(\dfrac{z}{\xi z_0} \right)^{n} , &  |z| \leqslant \xi z_0 \\
1 , & |z| > \xi z_0
\end{cases} , \\
\end{equation}
\begin{equation}
\delta(r,z) =
\begin{cases}
g(r,z) +\left[\delta_0- g(r,z) \right] \left(\dfrac{z}{\xi z_0} \right)^{n} , &  |z| \leqslant \xi z_0 \\
\delta_0 , & |z| > \xi z_0
\end{cases} ,
\end{equation}
in which $g(r,z) = N_m/[1+f(r,z)]$. Outside the IH region, the turbulence is believed to be CR-driven in principle.  Hence the diffusion is regarded as only rigidity dependent, namely $D_{xx} = D_0 \beta^\eta (\mathcal R/ \mathcal R_0)^{\delta_0}$.

% $D_{0}$ is adopted as 5.82$\times$$10^{28}$ cm$^{2}$s$^{-1}$ in this paper. 
% In the Conventional Propagation Model (CPM), $D_{xx}$ is simply correlated with particle rigidity, taken a form as $D_{0} \beta^{\eta} \left(\dfrac{R}{4GV} \right)^{\delta}$. The $\beta$ is for the particle velocity in units of light speed c. It should be noted that our new propagation model, 3D Spatial-Dependent Propagation Model (3DSDP), employs the same diffusion coefficients as that used in SDP model, but adopts the distinctive Galactic map of astrophysical source locations, which will be in exhaustively introduced in the next section.

\subsection{Spiral distribution of CR sources}
In the usual transport model, the CR sources are regarded as axisymmetric-distributed. This assumption is rational when the diffusion length of CRs is much larger than the characteristic distance between the adjacent spiral arms. However subject to the severe energy loss, the transport distance of the energetic electrons is much shorter. In this case, the specific position of the solar system and its neighbouring source distribution are expected to notably affect the observed spectrum of cosmic-ray electrons. Now it is widely accepted that our Galaxy is a typical spiral galaxy, in which the high-density gas inside spiral arms trigger the rapid star formation. So the cosmic-ray sources are highly concentrated in the spiral arms. There are still some uncertainties in the structure of the spiral arms, owing to our position in the Galaxy. While the outer part of the Milky Way seems to have four arms, the number of arms in the inner part is still being debated. The measurements for the spiral structure and number of spiral arms are reviewed in \citep{1995ApJ...454..119V, 1998ggs..book.....E, 2002ApJ...566..261V, 2017NewAR..79...49V}. The influences of spiral-distributed sources have been investigated in the recent studies \citep{2002PhRvL..89e1102S, 2009PhRvL.103k1302S, 2012JCAP...01..010B, 2012A&A...547A.120E, 2013PhRvL.111b1102G, 2014ApJ...782...34B, 2014NewA...30...32K, 2015APh....64...18W, 2015RAA....15...15L, 2015APh....70...39K, 2016ApJ...826...47B, 2017ApJ...846...67P, 2017MNRAS.466.3674N}.

%Despite our position in Galaxy,
% when evaluating cosmic ray fluxes at high energies where local sources could play a dominant role,may be important.
% in interstellar space
%of the solar system
% during the transport distance
%Subject to the location of the solar system in Galaxy, though the current observations  are still in some dispute, it is popular to accept
%in most published models
%, which is aimed at studying isolated radio pulsars

\begin{figure}[!htb]
\centering
\includegraphics[width=0.50\textwidth]{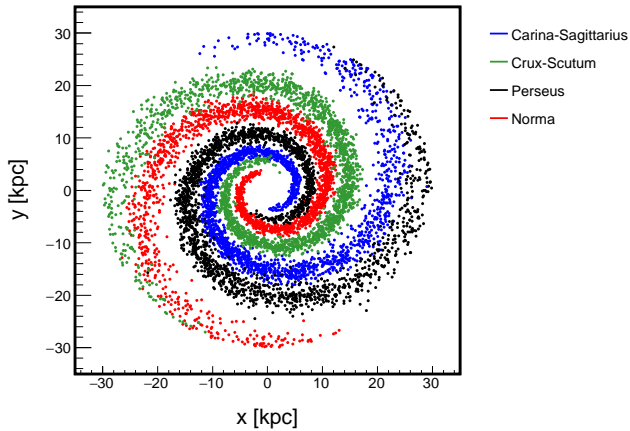}
\caption{Spiral distribution of CR sources in the Milky Way. The Galaxy is assumed to have four spiral arms, with the Sun lying between the Carina-Sagittarius and Perseus arms, about $8.5$ kpc away from the Galactic center.}
\label{spiral_dis}
\end{figure}

In this work, we adopt the spiral model established by \citep{2006ApJ...643..332F}. The whole Galaxy is considered to be made of four major arms spiraling outward from the Galactic center, as shown in Fig. \ref{spiral_dis}. For the $i$-th arm centroid, the locus can be expressed analytically by the logarithmic curve: $\theta(r) = k^{i} \ln(r/r^{i}_{0}) + \theta_{0}^i$, where $r$ is the distance to the Galactic center. Table \ref{spiral arms} lists the values of $k^{i}$, $r^{i}_{0}$ and $\theta_{0}^i$ for the each arm. Along the spiral arm, there is a spread in the radial coordinate that follows a normal distribution
\begin{equation}
f_i = \dfrac{1}{\sqrt{2\pi} \sigma} \exp \left(-\dfrac{(r-r_i)^2}{2\sigma^2} \right) ~, ~~~ i \in [1,2,3,4] ~,
\end{equation}
where $r_i$ is the inverse function of the locus of the $i$-th spiral arm and the standard deviation $\sigma$ is taken to be $0.07 r_i$. The density of CR sources at different radii is proposed to conform with the radial distribution under axisymmetric case. The solar system lies between the Carina-Sagittarius spiral centroid and Perseus spiral centroid. %$\langle z_{0} \rangle$, $100$ pc.

%In our 3DSDP model, we release the azimuthal symmetrical source distribution and develop this more realistic spiral pattern to describe the picture of our home Galaxy. By introducing this modification, the striking effects activating the propagation calculations of CRs, especially CREs, are in comparison with the results excepted in SDP and CMP models comprehensively.

  % develop SSDP model by some s of replacing the  CR source distribution with this
%    distribution of SNRs in the Milky Way,
%   The SSDP mentioned in our latter calculation of spectra for CRs and CREs are developed by the modification
%   taking account for the .

\begin{table*}[!t]
\begin{center}
\caption{\label{tab1}The values of the parameters $k^{i}$, $r^{i}_{0}$ and $\theta_{0}$ for four Galactic arm centroids\\}
 \label{spiral arms}
\begin{tabular}{|c|c|c|c|c|}
\hline
$i$-arm  &    name                &   $k^{i}$ (rad)     &    $r^{i}_{0}$ (kpc)  &     $\theta_{0}^i$ (rad)      \\\hline
1    &   Norma                     &   4.25                 &      3.48                   &     0                        \\
2    &   Carina-Sagittarius    &  4.25                 &       3.48                   &    4.71                  \\
3    &   Perseus                    &  4.89                  &      4.90                   &     4.09                  \\
4    &   Crux-Scutum          &  4.89                  &      4.90                   &   0.95                     \\
\hline
\end{tabular}
\end{center}
\end{table*}

\subsection{Local source}
%SNRs, microquasars
% in the Magnetic Halo
%within the distance and age equal to $1$ kpc and $10^{6}$ years respectively
%Because TeV CREs suffer severe energy losses via synchrotron and inverse Compton processes, it is indicated that the nearby sources, for example pulsars, pulsar wind nebula (PWNe), are able to contribute the excess of spectra of CREs above $100$ GeV. 
%could release the same account of $e^{\pm}$ pairs firstly confined in its nebula afterwards released continuously into the surrounding dense cloud for the finite duration of $\tau_{0}$ = 0.
%1$\times$$10^{6}$ years. 

To account for the excess of positron above $10$ GeV, we consider a pulsar nearby the solar system, which injects the electron and positron pairs instantaneously. The injection spectrum of this pulsar is assumed to be a power law:
\begin{equation}
Q(E)=Q_{0}\left(\dfrac{E}{1 ~\rm GeV}\right)^{-\gamma} ~.
%\left\{
%\begin{array}{lll}
%& \left( \dfrac{E}{1 ~\rm GeV}\right)^{-\gamma_{1}},  & {E \leq E_1} \\
%\\
%& \left( \dfrac{E_1}{1 ~\rm GeV} \right)^{-\gamma_1} \left( \dfrac{E}{E_1} \right)^{-\gamma_{2}} \exp \left(-\dfrac{E}{E_{\rm c}} \right),  & {E > E_1} \\
%\end{array}
%\right.
\label{eq:nearby}
\end{equation} 
%It must be emphasized that the calculated results devoted by this pulsar is defined as the nearby component, which differs with the background one from the CR sources distributed in the Milky Way.
%are adopted the values of 2.89 and 2.48, which are verified  offer the best fit to the data of $e^{\pm}$ spectra.

\section{Results}
%with the semi-analytic equations\ref{eq:transport},
%use the numerical package of DRAGON to solve three cases of the diffusion equations.
%which is the enhanced version of GALPROP by the improvements of the diffusion coefficients \citep{1998ApJ...509..212S}
% In this work, we expand DRAGON code to equip with one more alternative of the spiral arms distribution of CR sources.
%On this basis of the development, three kinds of various CR propagating situations (CMP, SDP, 3DSDP) are carefully discussed in the following content. 
%affecting spectra of CRs below tens of GeV

%The spectra of proton, electron and positron under the spatial-dependent propagation plus spiral distribution of sources are computed. We also compute the spectra under the conventional diffusion and the spatial-dependent propagation with axisymmetric distribution for comparison. The anisotropy under the model of spatial-dependent propagation plus spiral-distributed sources is also predicted. The numerical package of DRAGON \citep{2008JCAP...10..018E} is applied to solve the CR propagation equation. To explain the energy spectra less than tens of GeV, which are significantly affected by the solar modulation, the well-known force-field approximation \citep{1968ApJ...154.1011G} is introduced, with the modulation potential $\phi$ adjusted to fit low energy spectra.% well with an appropriate modulation potential $\phi$.

\subsection{B/C ratio and proton spectrum}
\begin{figure}[!htb]
\centering
\includegraphics[width=0.55\textwidth]{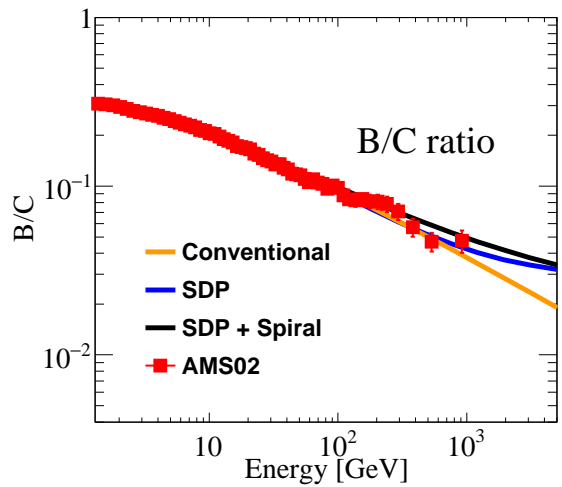}
\caption{The fitting to the B/C ratio under three propagation models, which are conventional(orange), SDP(blue) and SDP plus spiral-distributed sources(black) respectively. The data of AMS-02 experiment are taken from \citep{2015PhRvL.114q1103A}.
%The various model results come from the expectations of conventional model, SDP model and SDP + spiral distribution model, which correspond to the orange, blue and black solid line respectively.
}
\label{fitBCratio}
\end{figure}

%Firstly, we look at the fitting result of the spectrum for primary proton.
 %, in which the black dots and red squares represent the observations from the AMS02 experiment below TeV energy region and the observation from the CREAM experiment above TeV energy region.
% The green, blue and violet line correspond to the propagated spectra under the scenarios of CMP, SDP, 3DSDP models respectively. 
%(orange solid line) (blue solid line)  (black solid line)
%All of them obviously fit well with the observational points from AMS02 experiments represented as red dots.
%shown  of diffusion coefficients, solar modulation potentials and injection spectrum parameters
First of all, to determine the transport parameters, the ratio of B/C is fitted, which are illustrated in the Fig. \ref{fitBCratio}. The orange, blue and black solid lines correspond to the models of the conventional, SDP and SDP plus spiral-distributed sources respectively. The fitted transport parameters are listed in table \ref{tab:trans_prot_inj}. Compared with the conventional picture, the SDP scenarios expect that the B/C ratio could arise above TeV energies, which are clearly visible in the figure. 

\begin{figure}[!htb]
\centering
\includegraphics[width=0.55\textwidth]{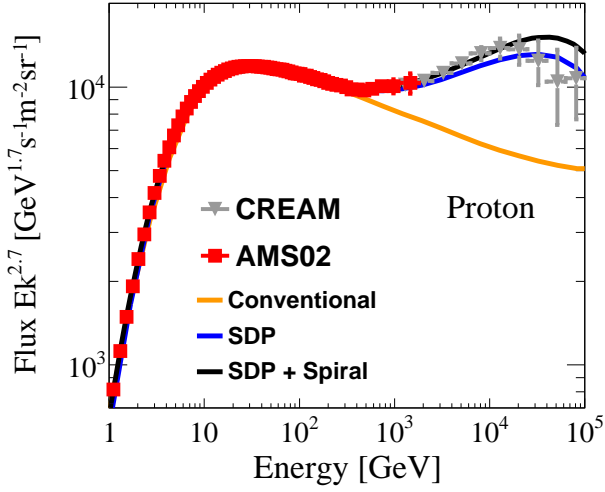}
\caption{The proton spectra under three propagation models, i.e.  conventional(orange), SDP(blue) and SDP plus spiral-distributed sources(black). The experiment data are taken from AMS-02 (red square) \citep{2015PhRvL.114q1103A,2017PhRvL.119y1101A} and CREAM (grey inverted triangle) \citep{2009ApJ...707..593A,2017ApJ...839....5Y}.}
    \label{proton_dragon}
\end{figure}

Fig. \ref{proton_dragon} shows the corresponding proton spectra. The red squares and gray inverted triangles are from the AMS-02 \citep{2015PhRvL.114q1103A} and CREAM experiments \citep{2017ApJ...839....5Y}. The conventional model predicts a single power-law above tens of GeV. In contrast, under SDP model, the propagation of energetic CRs is dominated by IH region, in which the rigidity dependence of the diffusion coefficient is flatter than outside. Thus in this case the proton spectra show a significant hardening above $\sim 200$ GeV, which could well reproduce the observations of both AMS-02 and CREAM experiments. Moreover, we could see that the distribution of sources make no difference to the proton flux at lower energies. Since the diffusion length of low energy protons is much longer than the distance between two neighbouring spiral arms, the distribution of CR sources does not prominently affect the spectrum. But for high energy protons, the major diffusion region is at the IH region, whose thickness is comparable to the distance between two neighbouring spiral arms, the source distribution could not be neglected. It can be seen that at TeV energies, the spiral distribution could produce a harder spectrum. The parameters of proton injection spectrum are listed in table \ref{tab:trans_prot_inj}.

%In the conventional diffusion, the propagated proton spectrum is of single power law, which could not explain the hardening of the primary proton. But under spatial-dependent propagation model, the propagated spectrum is a broken power law, in which above $\sim 200$ GeV, the spectrum gradually  becomes hardening. Thus it could naturally reproduce the hardening observed by both AMS-02 and CREAM experiments. Since the diffusion length of proton is much longer than the distance between two consecutive spiral arms, the distribution of CR sources does not prominently affect the spectrum. Less than  TeV, spiral and axisymmetric distributions are very close. Above tens of TeV, the proton under spiral distribution is harder.

%   It can be seen clearly the 3DSDP model has a priority to reproduce the hardening observed by both AMS02 and CREAM than the two other models.
%Noted that the proton fluxes expected by these propagation models are only taken accounts for the contribution from the background CR sources in the Galactic Disk.

\begin{table*}[h]
\begin{center}
\caption{The parameters for three different models including propagation and injection spectrum of proton}
\label{tab:trans_prot_inj}
\begin{tabular}{|c|c|c|c|}\hline
Parameters                                                                                   &   SDP+spiral                  &  SDP+axisymmetric              &  conventional         \\
\hline
$D_{0}$                                                                                      &   9.87                            &  4.6                                    &  5.82                    \\
$\delta_{0}$                                                                                &    0.65                           &  0.6                                     &  0.6                    \\
$z_{h}$[kpc]                                                                               &    6                                &   5                                       &  4                       \\
$N_{m}$                                                                                    &    0.27                            & 0.17                                     &                          \\
$v_{A}$ [km$\cdot$$s^{-1}$]                                                     &    6                                 &  6                                        &  30                  \\
\hline
 $q_{0}^{\rm p}[\rm cm^{-2}sr^{-1}s^{-1}GeV^{-1}]$              &   4.32$\times 10^{-2}$   &  4.36$\times 10^{-2}$         &  4.45$\times 10^{-2}$               \\
$\nu_{1}^{\rm p}$                                                                     &   2.0                               & 2.0                                      & 1.75                     \\      
$\nu^{\rm p}_{2}$                                                                     &   2.3                               & 2.4                                      & 2.25                  \\
${\cal R}_{\rm br}^{\rm p}$ [GV]                                                &   5.5                               & 5.5                                     & 9.9                 \\
$\phi^{\rm p}$[MV]                                                                     &   830                              &   830                                    &   560                  \\
\hline
\end{tabular}
\end{center}
\end{table*}

%\begin{center}
%\begin{table}[h]
%\begin{center}
%\caption{The parameters of injection spectrum for primary proton\\}
% \label{proton_spectrum}
%\begin{tabular}{|c|c|c|c|c|c|}\hline
%  &  $q_{0}^{\rm p}[\rm cm^{-2}sr^{-1}s^{-1}GeV^{-1}]$     &   $\nu_{1}^{\rm p}$     &  $\nu^{\rm p}_{2}$   &  ${\cal R}_{\rm br}^{\rm p}$ [GV]    &  ${\cal R}_{\rm c}$ [GV]     \\\cline{1-5}
%4.45$\times 10^{-2}$                                            &      2.0                     &  2.4                      &  5.5                     &  1.8 $\times 10^{5}$\\\hline
%\end{tabular}
%\end{center}
%\end{table}
%\end{center}

\subsection{Spectra of electron and positron}

%There are two kinds of major CRE sources above $100$ GeV. One is the primary electrons from the injection of the Galactic sources including the background astrophysical sources and the young nearby sources, and the other is produced by p-p collision during the transport of CRs in Galaxy, which is called the secondaries.
%, the positron spectrum 
%and the black line is the sum of them
%To well reproduce the excess of positron, the injection spectrum of local pulsar has to have a broken power law with an exponential cutoff.  Here we adopt  $E_{\rm c} as 70$ TeV.
%Besides that, it should be mentioned that the background component of $e^{+}$ must be enlarged to describe well the data points at low energy region.
%This has been mentioned in \cite{2015APh....60....1Y}.

\begin{figure*}[!htb]
\centering
\includegraphics[height=12.cm, angle=0]{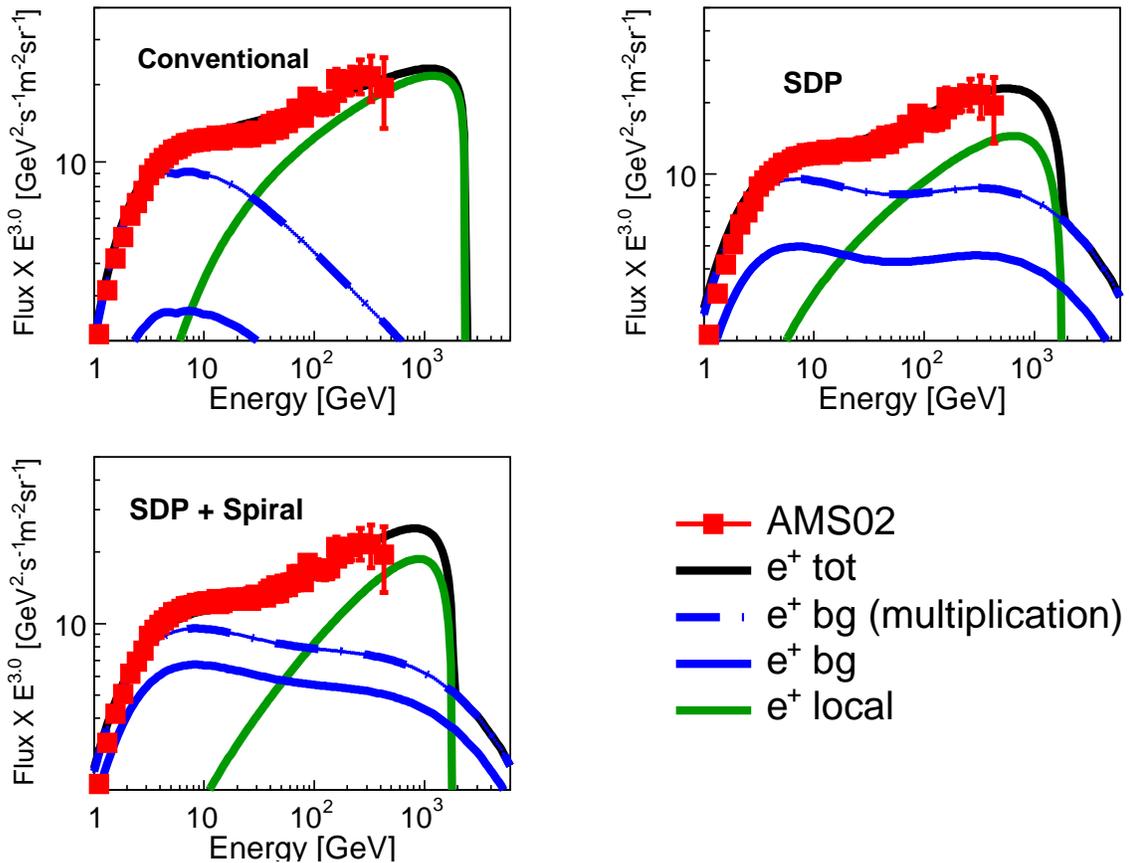}
\caption{The positron spectra computed under under three transport models, i.e. conventional (upper left), SDP (upper right), SDP + spiral distribution (lower left). The red squares are the measurements from the AMS-02 experiment \citep{2014PhRvL.113l1102A}. The blue and green solid lines represents the fluxes from the background and the local sources respectively. To well describe the positron flux, the background fluxes are multiplied by a factor $c^{e^+}$, which are shown as the dash-dot blue lines.}
\label{posi_detailed}
\end{figure*}

In Fig. \ref{posi_detailed}, we compare the positron spectra under three transport models. The red squares are from AMS-02 experiment \citep{2014PhRvL.113l1102A}. The blue and green solid lines are the contributions from background and local sources resepctively. Since the background positron flux is much lower, the observed positron data could not be described even with a local source. This could be caused by the possible uncertainties from the hadronic interactions, propagation models, the ISM density distributions, and the nuclear enhancement factor from heavy elements. In this work, the background positron flux is multiplied by a factor $c^{e^+}$, which is shown as a blue dash-dot lines. The black line is the sum of enhanced background and local fluxes. We could find that under the SDP model, the yield of second positrons is appreciably more than the conventional model. Therefore, the required enhancement factor $c^{e^+}$ in SDP model is only $1.92$ and $1.42$, smaller than the traditional model. Compared with axisymmetric case, $c^{e^+}$ under the spiral distribution is smaller. Meanwhile like proton, the propagated secondary positrons becomes hardening above tens of GeV. But due to the energy loss of positron during the transport, the broken energy shifts from $200$ GeV to $20$ GeV. Table \ref{electron_parameter} lists the distance and age of local source as well as the parameters for the corresponding injection spectrum.  

%compared with traditional model, $c^{e^+}$ under SDP model
% (black line)
%The enlarged coefficients, i.e. c, are respectively chosen to be 3.5, 1.92 and 1.42 corresponding to in conventional, SDP + axisymmetric and SDP + spiral model and enlarged background positron fluxes are demonstrated as the dashed blue lines in Fig. \ref{posi_detailed}.

\begin{figure*}[!htb]
\centering
\includegraphics[height=12.cm, angle=0]{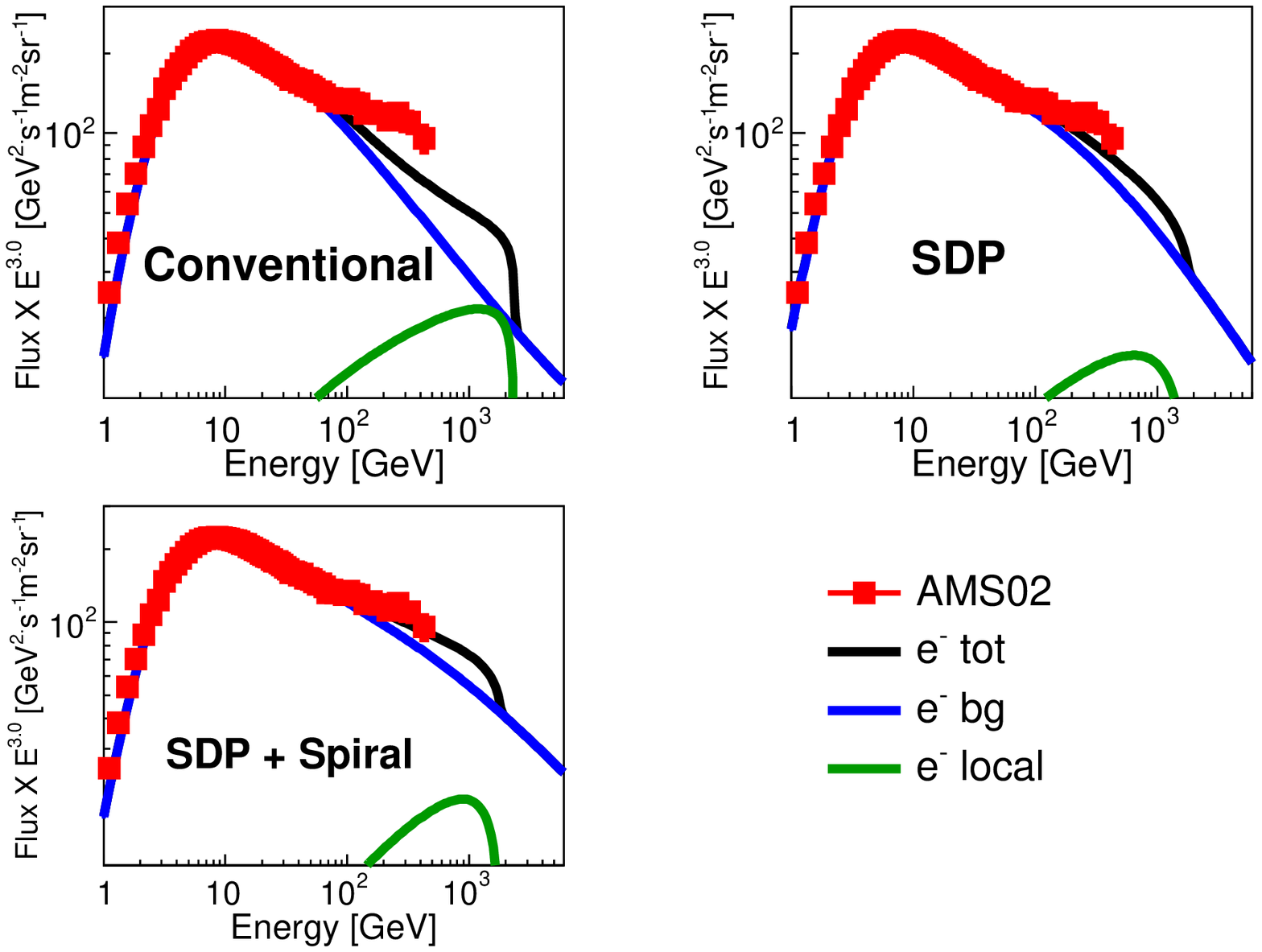}
\caption{The electron spectra computed under under three transport models, i.e. conventional (upper left), SDP (upper right), SDP + spiral distribution (lower left). The red squares are the measurements from the AMS-02 experiment \citep{2014PhRvL.113l1102A}. The blue and green solid lines represents the fluxes from the background and the local sources respectively. }
\label{elec_detailed}
\end{figure*}

Fig.\ref{elec_detailed} shows the corresponding electron spectra under three transport models. The red squares are the AMS-02 experiment \citep{2014PhRvL.113l1102A}. The blue and green solid lines are the contributions from background and local sources resepctively, while the black line is the sum of them. It can be seen that under conventional model, the computed electron flux could not explain the observational data, even with only a local source. It can be inferred that an additional electron component is needed in the case of the conventional transport scenario \citep{2017PhRvD..96b3006L}. This is also indicated by the analysis of \cite{2015PhLB..749..267L}. But under the SDP model, the calculated electron flux conforms with the observation much better. This is because that the propagated electron spectra has a hardening above $\sim 10$ GeV, which elevates the background electron flux. With spiral-distributed sources, the electron flux could well describe the observed electron flux. The parameters of injection spectrum of primary electrons from background and local sources for three propagation models are also given in Table \ref{electron_parameter}.
%The exponential cutoff is chosen to be $E_{\rm c} = 70$ TeV.  

\begin{figure}[!htb]
\centering
\includegraphics[width=0.5\textwidth]{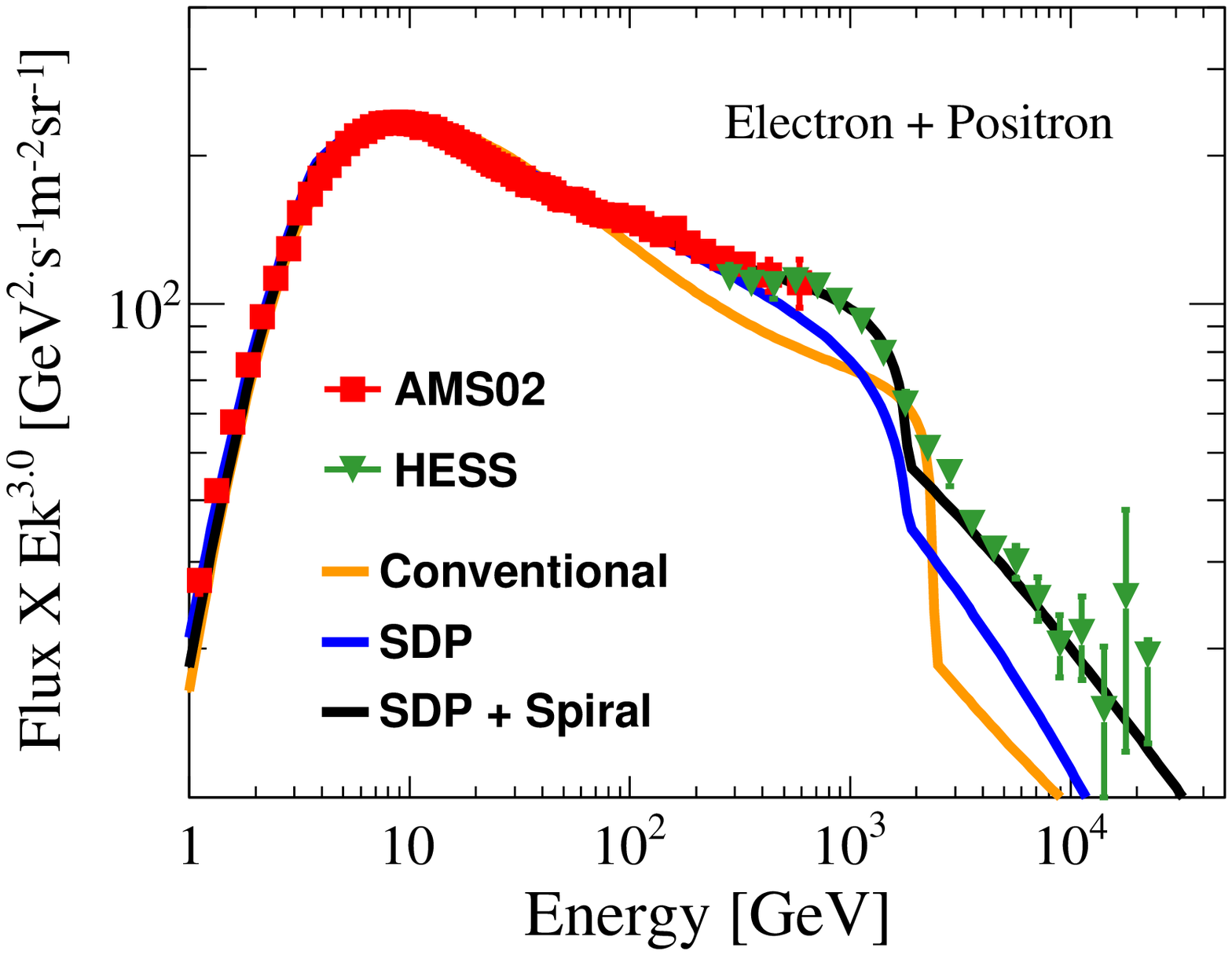}
\caption{The total electron spectra (electron + positron) computed by three transport models, i.e. conventional (orange), SDP (blue), SDP + spiral distribution (black). The red squares and green inverted triangle are the measurements from the AMS-02 \citep{2014PhRvL.113v1102A} and HESS experiments respectively.}
\label{CREs_spectrum_ams02}
\end{figure}

Fig. \ref{CREs_spectrum_ams02} further shows the total electron spectra computed by different propagation models in comparison with AMS-02 \citep{2014PhRvL.113v1102A} and HESS experiments. The orange, blue and  black solid lines correspond to the conventional, SDP and SDP with spiral distribution respectively. In the conventional model, the obtained background electron flux is a single power-law, which is not enough for the high-energy electron component. For the SDP model, the traditional distribution of sources could marginally explain both electron and positron spectra less then TeV energies. But when considering the latest observations of HESS, it is still not enough. By introducing a spiral distribution, the high energy electron flux has been significantly boosted. We could see that the SDP plus spiral distribution can well reproduce total electron spectrum within the whole energy range, compared with the other two models. It is noteworthy that there is a sharp break around several TeV for each model, which are brought about by the plummet of local flux. This is caused by our simplified cooling time of electron, which is fixed to $10^{16}$s. When considering the distribution of background photons, the sharp break is expected to disappear. 

\begin{table*}[h]
\begin{center}
\caption{The parameters of injection spectrum for background electron and local electron and positron}
 \label{electron_parameter}
\begin{tabular}{|c|c|c|c|}\hline
parameters                       &  SDP+spiral               & SDP+axisymmetric            & conventional       \\
\hline
$q_{0}^{\rm e^-}[\rm cm^{-2}sr^{-1}s^{-1}GeV^{-1}]$     &   0.30   & 0.30  &  0.16  \\
$\nu_{1}^{\rm e^-}$              &   1.14               & 1.53                      & 1.69  \\
$\nu^{e^-}_{2}$                    &   2.72               & 2.89                      & 2.81  \\
${\cal R}_{\rm br}^{\rm e^-}$  [GV]    &   4.6     & 4.6   & 4 \\
\hline
$r$ [kpc]                & 0.25                          & 0.21                              & 0.21 \\
$t$ [yr]               &  1.7$\times 10^{5}$  & 1.7$\times 10^{5}$        & 1.3$\times 10^{5}$ \\
$Q_{0}$ [GeV$^{-1}$]            & 1.9$\times 10^{50}$ & 2.7$\times 10^{50}$      & 2.7$\times 10^{50}$ \\
$\gamma$             &  2.46                         & 2.60                               & 1.84 \\
$c^{e^+}$                         & 1.42                          & 1.92                                & 3.5  \\
$\phi^{\rm e^{+}}$[MV]          & 650                           & 690                                 &1300 \\
$\phi^{\rm e^{-}}$ [MV]          & 650                           & 650                                 &1250 \\
\hline
\end{tabular}
\end{center}
\end{table*}

\subsection{Anisotropy of electron}
We further calculate the dipole anisotropies of electron as a function of energy under these propagation models. The dipole anisotropy is usually defined as
\begin{equation}
\delta = \frac{3D_{xx} }{v} \dfrac{|\nabla \psi|}{\psi} ~.
\end{equation}
Due to the distribution of Galactic SNRs with a higher density at the inner Galactic disk, there is inevitably a radial gradient of CR density from the direction of the Galactic center. Thereupon in the scenario of steady-state propagation, the anisotropy scales with the diffusion coefficient so that it grows with the energy. 

\begin{figure*}[!htb]
\centering
\includegraphics[height=7.cm, angle=0]{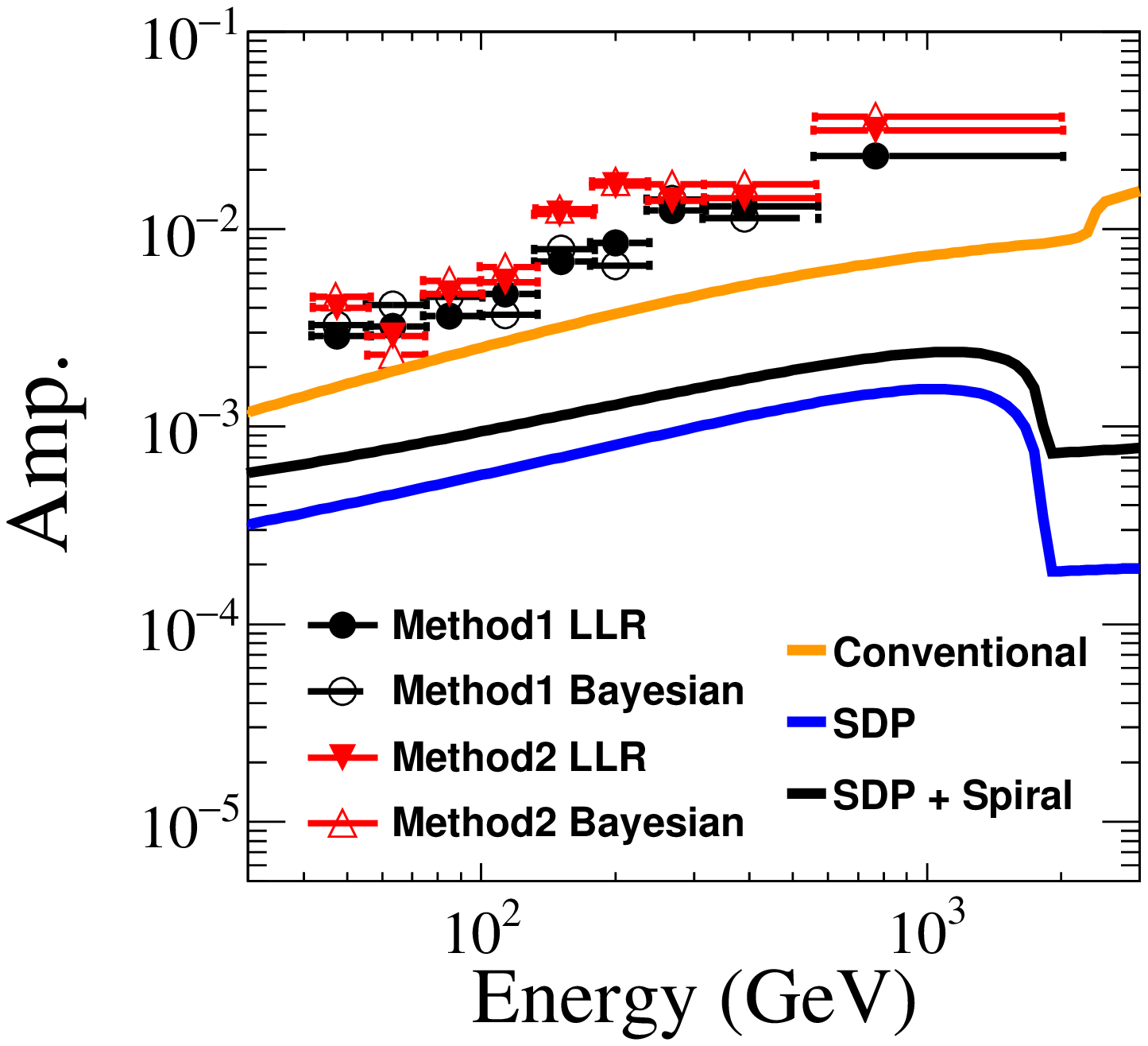}
\caption{Anisotropies of electron with the local source at the direction of Galactic center. The orange, blue and black solid lines show the anisotropies under the conventional, SDP and SDP+spiral distribution respectively. The upper limits are from the Fermi-LAT experiment \citep{2017PhRvL.118i1103A}, which are shown as the black and red markers.}
\label{anisotropy_draw_spri}
\end{figure*}

Fig.\ref{anisotropy_draw_spri} illustrates the anisotropies of electron when the local source is in the direction of Galactic center. The orange, blue and black solid lines show the anisotropies  under the conventional, SDP and SDP+spiral distribution respectively. In the conventional model, due to a larger diffusion coefficient, the expected background anisotropy is obviously large. When the local source is at the direction of Galactic center, its influence to the total anisotropy is inconspicuous, while the background has an overwhelming contribution, as shown in the left figure. In this case, the total anisotropy is close to the observed upper limit by the Fermi-LAT experiment. 

Different from the conventional model, the background anisotropy under SDP model is well below the current upper limit. This is due to the slower diffusion coefficient around the Galactic disk. Compared with the axisymmetric case, the spiral distribution induces a larger background anisotropy. From the figure, it can be seen that the anisotropy from the local source is much larger between $100$ and $1000$ GeV. Even so, the total anisotropy is still much smaller. The current and future experiments, which observe the electrons, for example DAMPE and LHAASO, could test our model.

\section{Conclusion}
% Galactic CR sources distributing along the realistic 
%, which replaces the previous axisymmetric distribution

With the development of instruments, the detections of cosmic-ray electron and positron have been greatly improved in these recent years. The precise measurements unveil some interesting features and raise the new challenges for the traditional propagation model. %The traditional propagation model can hardly explain the current observations.
%As the expected spectrum of energetic CREs performs softer due to severe energy losses,
%propagation under with spiral distribution of CR sources. 

In this work, we apply the SDP + local source model to study both electron and positron spectra. We also introduce the spiral model to account for more realistic distribution of CR sources. We find that even with one local source, the traditional propagation model could not self-consistently describe both electron and positron spectra, thus fail to reproduce the total electron spectrum. However, compared with the conventional model, both positron and electron spectra have a spectral hardening above tens of GeV in the SDP model, which elevates the background flux. Meanwhile, the high energy electrons have shorter diffusion length, the distribution of background sources have non-negligible influence above TeV energies. Taking into account of the spiral distribution, the TeV break of the total electron spectra could be well reproduced by the SDP model, which conforms with the data up to $25$ TeV.

Furthermore, we compute the anisotropy of electron. In the conventional model, the background anisotropy is larger, which is very close to the latest upper limit set by the Fermi-LAT experiment. But the SDP model predicts a much lower anisotropy by background sources, which is due to the smaller diffusion coefficient around the Galactic disk. In this case, the local source could have significant influence above $\sim 100$ GeV. But even including the local source, the total anisotropy is still very small. We hope the experiments, for example DAMPE and LHAASO, could test out model.

\section*{Acknowledgements}
This work is supported by the National Key Research and Development Program of China (No. 2016YFA0400200), the National Natural Science Foundation of China (Nos. 11875264, 11635011, 11761141001, 11663006).

\bibliographystyle{apj}
\bibliography{ref}

\end{document}